\documentclass[final,aps,prl,twocolumn,superscriptaddress,showpacs]{revtex4-1}
\usepackage{graphicx}% Include figure files
\usepackage{dcolumn}% Align table columns on decimal point
\usepackage{amsmath,bm}% bold math
\usepackage{color}%
\usepackage{ulem}

\newcommand{\beginsupplement}{%
        \setcounter{table}{0}
        \renewcommand{\thetable}{S\arabic{table}}%
        \setcounter{figure}{0}
        \renewcommand{\thefigure}{S\arabic{figure}}%
     }
\begin{document}

%\listoftodos
%\newpage

\title{Symmetry dictated grain boundary state in a two-dimensional topological insulator}

\author{Hyo Won Kim}
\email[E-mail: ]{hyowon98.kim@samsung.com}
\affiliation{Samsung Advanced Institute of Technology, Suwon 13595, Korea}

\author{Seoung-Hun Kang}
\affiliation{Korea Institute for Advanced Study, Seoul 02455, Korea}

\author{Hyun-Jung Kim}
\affiliation{Korea Institute for Advanced Study, Seoul 02455, Korea}

\author{Kisung Chae}
\affiliation{Korea Institute for Advanced Study, Seoul 02455, Korea}

\author{Suyeon Cho}
\affiliation{Division of Chemical Engineering and Materials Science, Ewha Womans University, Seoul, 03760, Korea}

\author{Wonhee Ko}
\affiliation{Samsung Advanced Institute of Technology, Suwon 13595, Korea}
\affiliation{Center for Nanophase Materials Sciences, Oak Ridge National Laboratory, Oak Ridge, Tennessee 37831, USA}

\author{Se Hwang Kang}
\affiliation{Department of Energy Science, Sungkyunkwan University, Suwon, 440-746, Korea}

\author{Heejun Yang}
\affiliation{Department of Energy Science, Sungkyunkwan University, Suwon, 440-746, Korea}

\author{Sung Wng Kim}
\affiliation{Department of Energy Science, Sungkyunkwan University, Suwon, 440-746, Korea}

\author{Seongjun Park}
\affiliation{Samsung Advanced Institute of Technology, Suwon 13595, Korea}

\author{Sung Woo Hwang}
\affiliation{Samsung Advanced Institute of Technology, Suwon 13595, Korea}

\author{Young-Kyun Kwon}
\affiliation{Korea Institute for Advanced Study, Seoul 02455, Korea}
\affiliation{Department of Physics and
             Research Institute for Basic Sciences,
             Kyung Hee University, Seoul, 02447, Korea}

\author{Young-Woo Son}
\email[E-mail: ]{hand@kias.re.kr}
\affiliation{Korea Institute for Advanced Study, Seoul 02455, Korea}
%---------------------------------------------------------------------

\begin{abstract}
Structural imperfections such as grain boundaries (GBs) and dislocations are ubiquitous in solids and have been of central importance in understanding nature of polycrystals. In addition to their classical roles, advent of topological insulators (TIs) offers a chance to realize distinct topological states bound to them. Although dislocation inside three-dimensional TIs is one of the prime candidates to look for, its direct detection and characterization are challenging. Instead, in two-dimensional (2D) TIs, their creations and measurements are easier and, moreover, topological states at the GBs or dislocations intimately connect to their lattice symmetry. However, such roles of crystalline symmetries of GBs in 2D TIs have not been clearly measured yet. Here, we present the first direct evidence of a symmetry enforced Dirac type metallic state along a GB in 1T'-MoTe$_2$, a prototypical 2D TI. Using scanning tunneling microscope, we show a metallic state along a grain boundary with non-symmorphic lattice symmetry and its absence along the other boundary with symmorphic one. Our large scale atomistic simulations demonstrate hourglass like nodal-line semimetallic in-gap states for the former while the gap-opening for the latter, explaining our observation very well. The protected metallic state tightly linked to its crystal symmetry demonstrated here can be used to create stable metallic nanowire inside an insulator. 
\end{abstract}
%---------------------------------------------------------------------

% \keywords{graphyne, density functional theory, electronic structure}


\maketitle

%---------------------------------------------------------------------
% If in two-column mode, this environment will change to single-column
% format so that long equations can be displayed. Use sparingly.
%\begin{widetext}
%\end{widetext}

Grain boundary (GB) or dislocations are important structural imperfections in studying nature of polycrystals~\cite{sutton1995interfaces}. Besides their classical properties, nowadays, various interesting topological states along the GBs or dislocations can be realized inside topological insulators (TIs)~\cite{{kane2005quantum},{bernevig2006quantum},{konig2007quantum},{ran2009onedimensional},{drozdov2014onedimensional},{slager2014interplay}}, although their direct detection and characterization are challenging~\cite{{ran2009onedimensional},{slager2014interplay}}. In two-dimensional (2D) TIs, interesting topological properties of GBs~\cite{{slager2014interplay},{juricic2012}} can be detected more easily using local probes. Among many two-dimensional (2D) materials, various grain boundaries with distinct spatial symmetries can be realized in a 2D transition metal dichalcogenide, 1T'-MoTe$_2$ thanks to its special crystal structure. Because the 1T' phase of MoTe$_{2}$ can be understood as a static lattice distortion from its more symmetric but unstable 1T phase~\cite{keum2015bandgap}, there are chances to create disparate structural phase boundaries between crystalline domains with different orientations while growing the 1T' structure (hereafter we will drop 1T' for convenience). Our study of one-dimensional (1D) topological metallic states along GBs of MoTe$_2$ was based on scanning tunneling microscopy (STM) and spectroscopy (STS). In as-grown MoTe$_2$ sample surfaces, we observed an unprecedented GB structure satisfying a non-symmorphic glide-reflection symmetry and topological metallic state associated with it. Quite contrary to this, we could distinguish a semiconducting boundary state along the other boundary with a symmorphic mirror symmetry. Our measurement of existence of topological states depending on the symmetries of GBs was further confirmed by large scale atomistic simulations.

We first identify the topological nature of the surface layer of an as-grown MoTe$_2$ bulk sample. Although we measured tunneling spectra on the surface of the bulk sample, our measurements clearly indicate that the topmost layer behaves as a single layer detached from the rest of sample. Such a behavior has already been observed in a similar material~\cite{peng2017observation} as well as other layered crystals, e.g., graphene-like behavior of topmost layer of graphite~\cite{li2007observation} and Bi surface layer~\cite{drozdov2014onedimensional}. In our sample with GBs, the presence of GBs in the topmost layer promotes the lattice mismatch between the topmost layer and the others, further ensuring the single layer spectrum in the topmost one. Our simulated STM images and STSs also support the single-layer-like behavior of the topmost layer of the bulk MoTe$_2$ sample. The simulated images of the single layer MoTe$_2$ with varying applied voltage also agree with our observations very well (Supplementary Section I). Thanks to the effective isolation, we were able to measure the topologically protected metallic state along the truncated step edge (Supplementary Section II) by its dI/dV spectra and STM images agree well with previous studies~\cite{{peng2017observation},{jia2017},{tang2017quantum}}. 

% entire page. There is no need to do explicit centering.
\begin{figure}
\includegraphics[width=1.0\columnwidth]{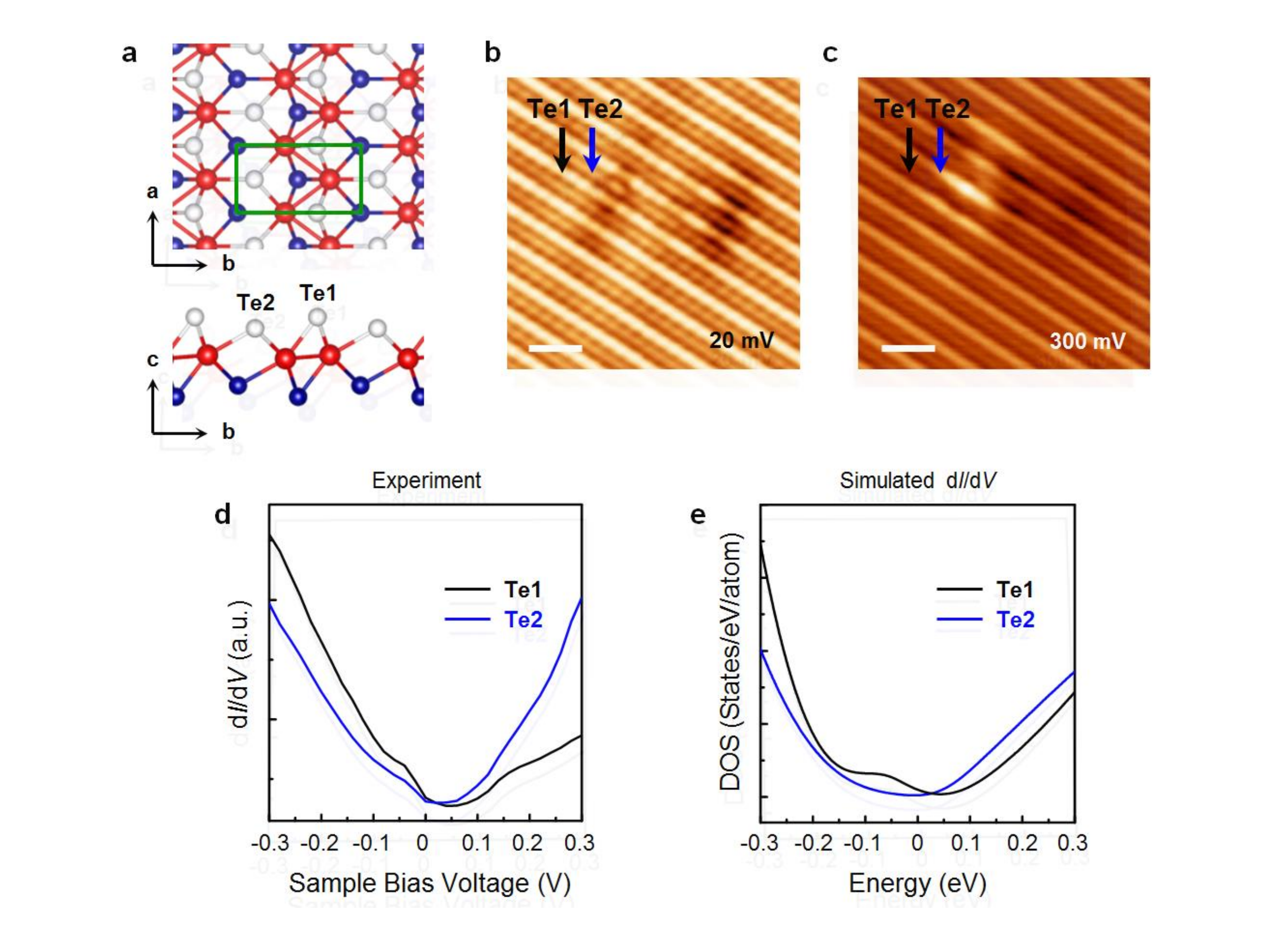}
\caption{(Color online) Morphology and electronic structure of a 1T'-MoTe$_2$. {\bf a,} Schematic illustration of 1T'-MoTe$_2$. The green box represents a 1T'-MoTe$_2$ unit cell containing Te1 and Te2 with a height difference relative to the transition metal plane. {\bf b, c,} STM topograph of the 1T'-MoTe$_2$ surface near a defect at Vs = 20~mV and Vs = 300~mV, respectively (It = 0.5~nA). Scale bars are 1~nm. The contrast between Te1 and Te2 in the STM image depends on the applied voltages. {\bf d,e,} experimentally obtained and DFT-calculated dI/dV spectra at Te rows, Te1 (black line) and Te2 (blue line), respectively. 
\label{Fig1}}
\end{figure}
%---------------------------------------------------------------------

 In MoTe$_2$, the chalcogen atoms (Te atoms) form quasi-one-dimensional parallel chains where adjacent chains (denoted by Te1 and Te2 rows, respectively) have alternative height variation with respect to the transition metal plane as shown in Fig.~\ref{Fig1}a. The inequivalent rows of Te atoms will play an important role in determining the local crystalline symmetry along GBs later. The height alternation spots in our STM images with different contrast. We found that this contrast difference between the two chains also depends on the applied bias voltages; for example, the contrast at 300~mV is reversed in Fig.~\ref{Fig1}c comparing to that at 20~mV in Fig.~\ref{Fig1}b (A defect is used as a marker to identify Te1 and Te2 rows in Figs.~\ref{Fig1}b and c.) (See details in Supplementary Section III). To figure out the effect of the electronic state, we measured the local dI/dV spectra of Te1 and Te2 rows. As is matched very well with images, the values of dI/dV at different rows cross each other at 20~mV in Fig.~\ref{Fig1}d. The different energy levels of Te1 and Te2 p-orbitals from our first-principles simulation can explain alternating contributions from two chalcogen atoms in determining crystal structures of MoTe$_2$ precisely (Supplementary Section III).

% entire page. There is no need to do explicit centering.
\begin{figure}
\includegraphics[width=1.0\columnwidth]{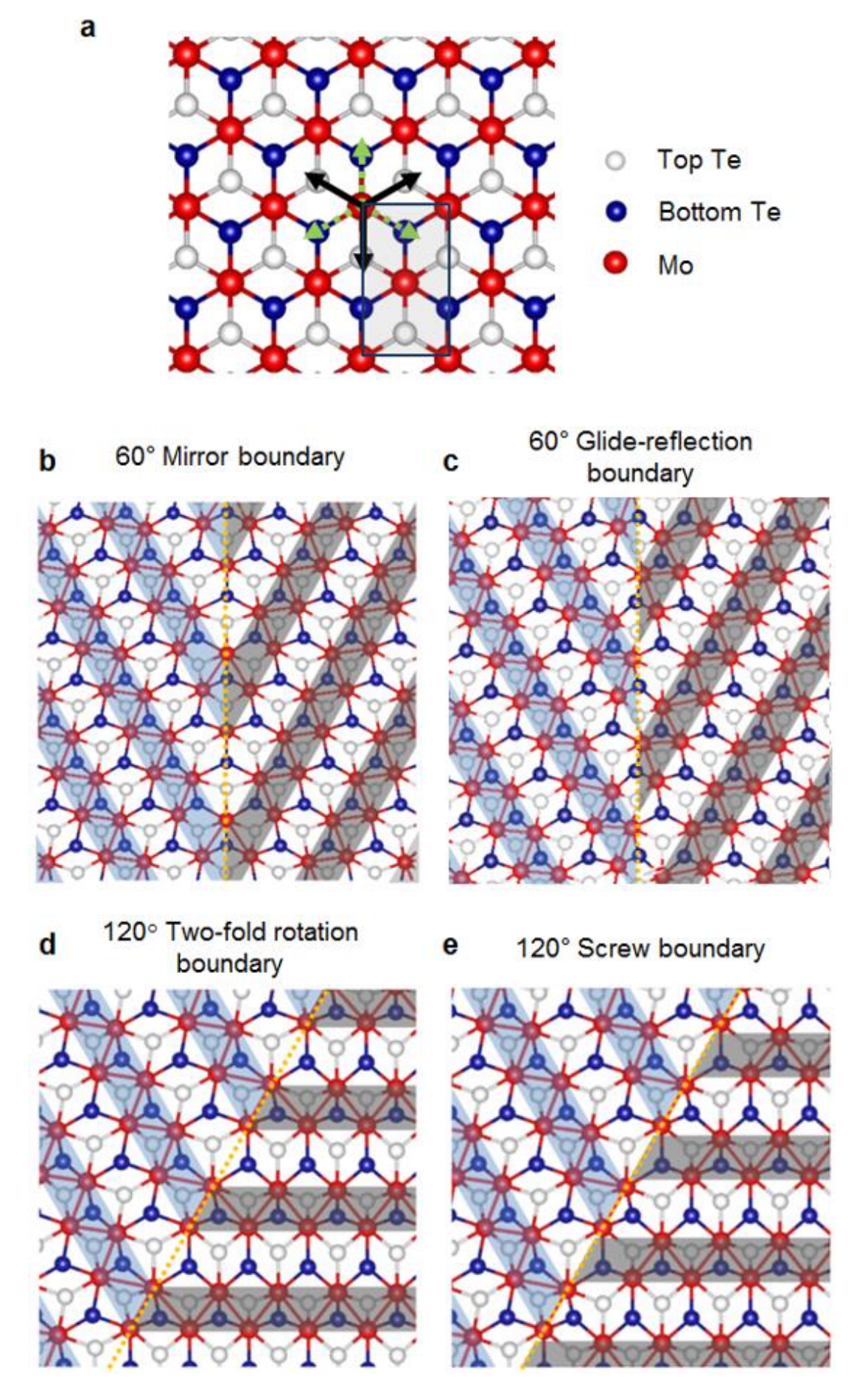}
\caption{(Color online) Possible grain boundaries from six orientation variants of 1T'-MoTe$_2$. {\bf a,} Three symmetry-equivalent directions (black arrows) of structural distortion in the 1T structure and their opposite three directions (green dotted arrows), which are distinguished by the directions towards top and bottom Te atoms, respectively. The rectangular is a primitive unit cell of the 1T' phase, which corresponds to the $1{\times}\sqrt{3}$ supercell of the 1T phase. {\bf b--e,} Possible grain boundary structures determined by the direction of structural distortion with respect to the boundary. To help guide the eyes, dimerized Mo atoms are shaded in blue and black colors.
\label{Fig2}}
\end{figure}
%---------------------------------------------------------------------

% entire page. There is no need to do explicit centering.
\begin{figure}
\includegraphics[width=1.0\columnwidth]{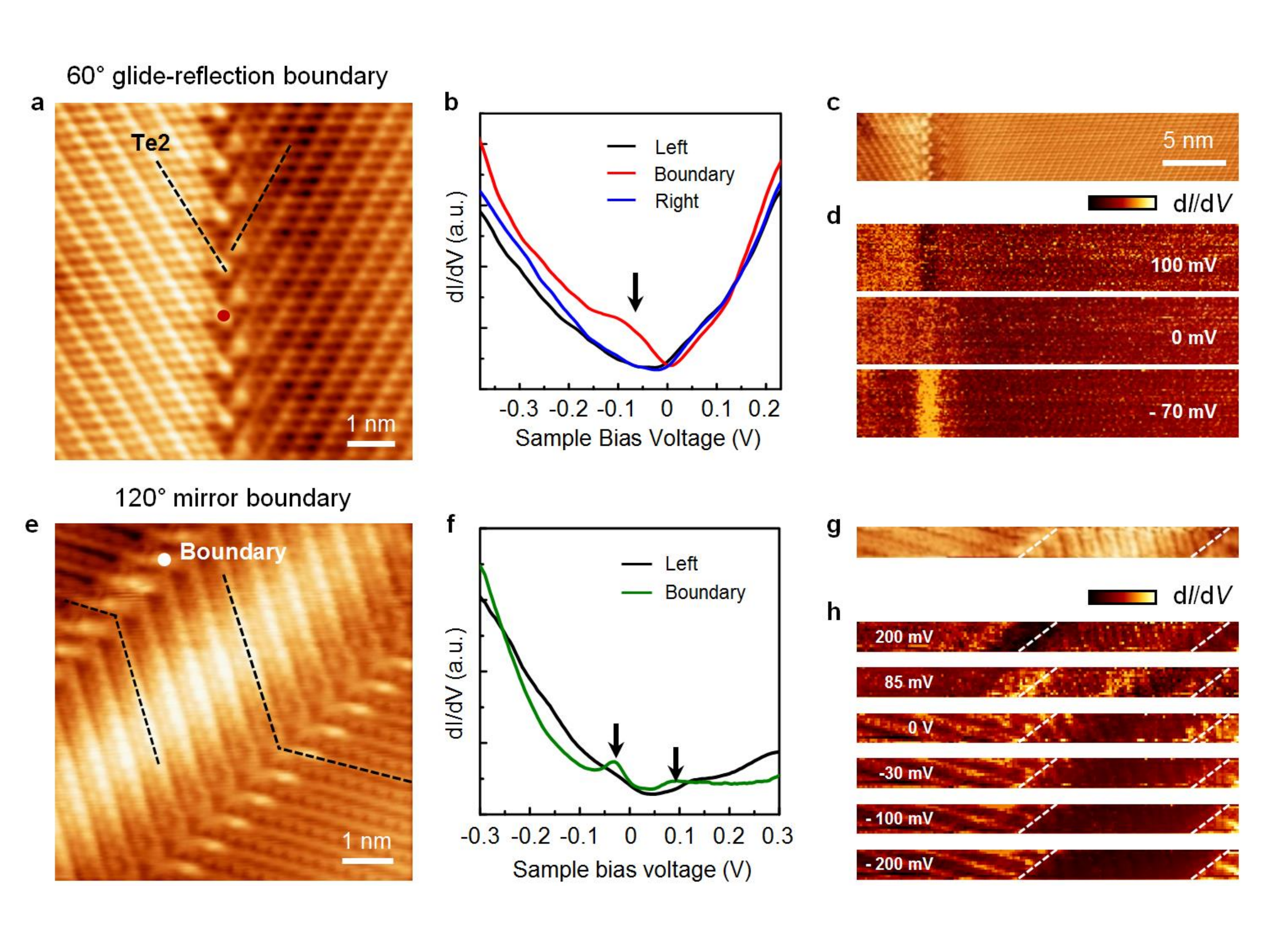}
\caption{(Color online) STM topograph and dI/dV spectra of 60$^\circ$ glide reflection and 120$^\circ$ two-fold rotation boundaries of 1T'-MoTe$_2$. {\bf a,} STM topograph of 60$^\circ$ glide reflection boundary (Vs = 0.3~V, I = 0.5~nA). {\bf b,} Averaged dI/dV spectra taken at left area (black line), and right area (blue line) and dI/dV spectrum obtained at the position indicated by the red dot in a (red line). {\bf c,} STM topograph. {\bf d,} dI/dV maps obtained over the area shown in {\bf c} for bias voltages V = 100, 0 and $-$70~mV from top to bottom. {\bf e,} STM topograph of the 120$^\circ$ two-fold rotation boundary (Vs = 0.3~V, I = 0.1~nA). {\bf f,} Averaged dI/dV spectrum taken at left area (black line), and dI/dV spectrum obtained at the position indicated by the white dot in {\bf e} (green line). {\bf g,} STM topograph, {\bf h,} dI/dV maps obtained over the area shown in {\bf g} for bias voltages V = 200, 85, 0, $-$30, $-$100 and $-$200~mV from top to bottom. The dashed white lines are used as guides indicating 120$^\circ$ boundaries. \label{Fig3}}
\end{figure}
%---------------------------------------------------------------------

Having established a single-layer-like behavior of the topmost layer, now we turn to formation mechanism of GBs. Suppose that a structural phase transition of a matrix from the high symmetry (1T) phase to the low one (1T') through dimerization of adjacent Mo rows occurs in multiple locations. There are six orientation variants, resulted from the six-fold improper rotational symmetry (see black and grey dotted arrows in Fig.~\ref{Fig2}a). Excluding trivial and high energy structures, the interfaces between two 1T' phase crystals in different orientations can make four inequivalent symmetric tilt GBs with the angles of either 60$^\circ$ or 120$^\circ$, which can be categorized by the GB operations that map the right side onto the left of crystal structure. We identified that those GB operations are 60$^\circ$ mirror, 60$^\circ$ glide-reflection, 120$^\circ$ two-fold rotation and 120$^\circ$ screw, and the corresponding boundary structures are shown in Figs.~\ref{Fig2}b-e. Those low energy GBs were found by a simple GB model considering point group symmetry in addition to the coincidence site lattice theory (Supplementary Section IV). Furthermore, we found that 60$^\circ$ glide-reflection and 120$^\circ$ two-fold rotation boundaries are energetically favorable ones because their strains are smaller than the others. We confirmed that those two structures match the experimentally observed ones (Figs.~\ref{Fig3}a, e) and are approximately 110~meV and 570~meV more stable than the other structures, respectively. We note that the 120$^\circ$ two-fold rotation boundary was theoretically predicted as a twin boundary induced by strain in single crystalline 1T'-MoTe$_2$~\cite{li2016ferroelasticity}, which was also observed experimentally~\cite{{naylor2016monolayer},{wang2019et}} while the 60$^\circ$ glide-reflection boundary has never been reported yet.

 At the 60$^\circ$ glide-reflection boundary, the STM image in Fig.~\ref{Fig3}a shows its non-symmorphic symmetry nature clearly. The bright rows of chalcogen atoms in the right side of the boundary show abrupt discontinuities to the left side ones. The bright rows in the right side can map onto the left ones by simultaneous operations of mirror and a half-unit translation with respect to the GB. The measured dI/dV spectrum at the boundary in Fig.~\ref{Fig3}b (red line) reveals a peak near $-$70~mV whereas the local dI/dV spectrum far away from it (black and blue lines) shows no feature at $-$70~mV similar with the local spectrum of the pristine sample (Fig.~\ref{Fig1}d). We note that the observed spectrum in Fig.~\ref{Fig3}b is very similar with recently reported topological edge states in 1T'-WTe$_2$~\cite{{peng2017observation},{jia2017},{tang2017quantum}} and our results on the step edge of 1T'-MoTe$_2$ (Supplementary Section II). The peak is spatially localized at the boundary within ~ 2~nm width as clearly seen in the spatial dI/dV maps in Fig.~\ref{Fig3}d and Supplementary Section II. 

The spatially and energetically localized one-dimensional metallic state along the GB with non-symmorphic symmetry is in sharp contrast to states in the other boundary with symmorphic symmetry. Unlike the case of the 60$^\circ$ glide-reflection boundary, the STM image of the 120$^\circ$ two-fold rotation boundary shows the continuous bright rows across the boundary with kinks as shown in Fig.~\ref{Fig3}e. So, the right side STM image can map onto the left one by two-fold rotation operation. Its dI/dV spectrum in Fig.~\ref{Fig3}f also shows the stark differences: two peaks at 85 and -30~mV (green line that locate away from the energetic position of the gap at $-$70~mV (black line)). The spatial distributions of the peaks at 85 and $-$30~mV in Fig.~\ref{Fig3}h indicate that the states associated with the peaks are broadly distributed along the boundaries and that their intensities are quite lower than that of the 60$^\circ$ glide-reflection boundary (further dI/dV maps are in Supplementary Section V).

% entire page. There is no need to do explicit centering.
\begin{figure}
\includegraphics[width=1.0\columnwidth]{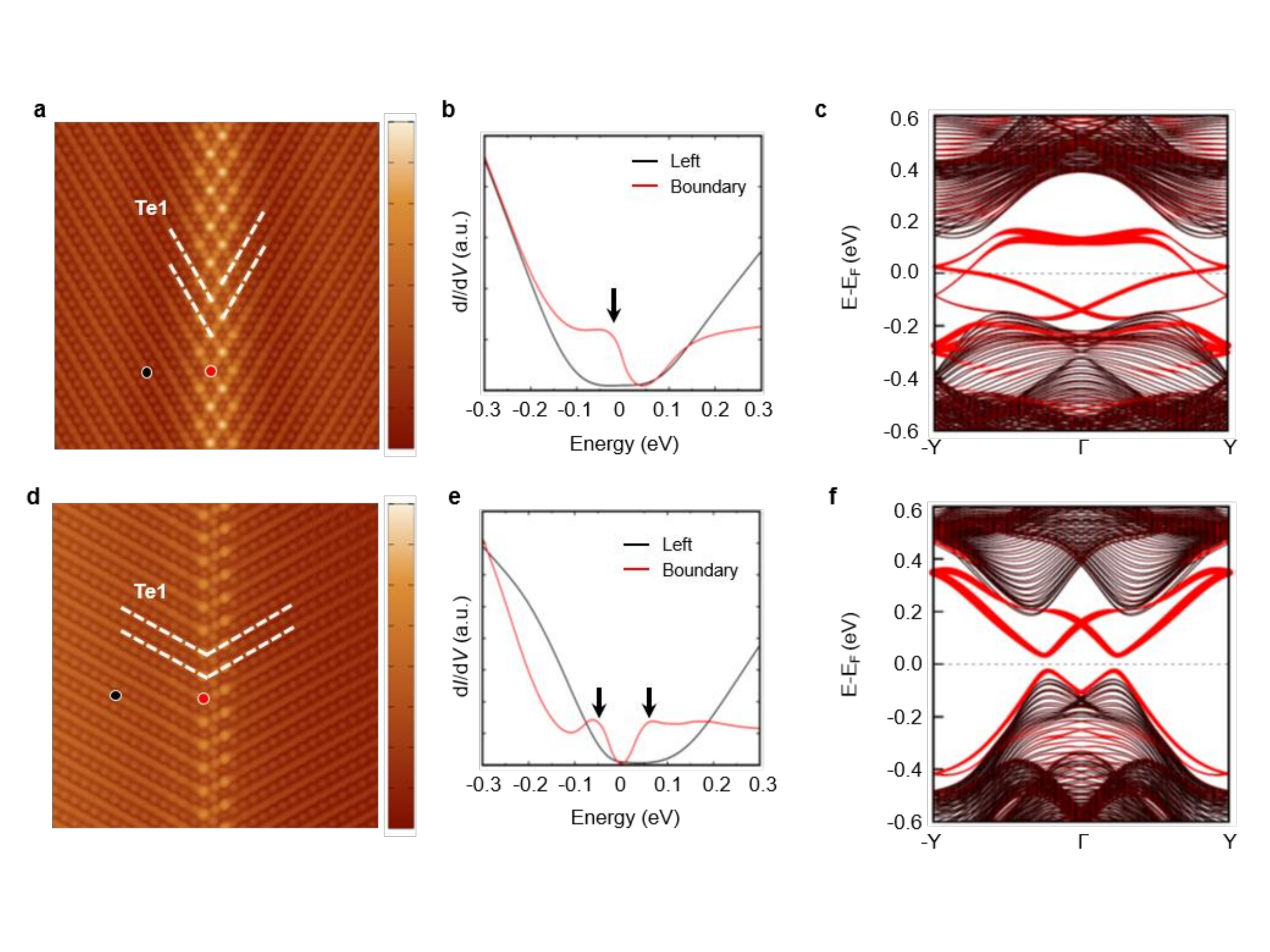}
\caption{(Color online) Theoretical electronic structures of 60$^\circ$ glide-reflection and 120$^\circ$ two-fold rotation boundaries. {\bf a,} Simulated STM image of 60$^\circ$ glide reflection boundary on the Vs = 0.3~V. {\bf b,} Simulated dI/dV at left area (black line) and boundary (red line) denoted by black and red dots in a, respectively. {\bf c,} Band structure for the 60$^\circ$ glide reflection boundary with twin boundary structure. {\bf d,} Simulated STM image of 120$^\circ$ two-fold rotation boundary on the Vs = 0.3~V. e, Simulated dI/dV at left area (black line) and boundary (red line) denoted by black and red dots in d, respectively. {\bf f,} Band structure for the 120$^\circ$ two-fold rotation boundary with twin boundary structure. \label{Fig4}}
\end{figure}
%---------------------------------------------------------------------

The band crossing in our calculated band structures in Fig.~\ref{Fig4}c. clearly explains the origin of the metallic states in the 60$^\circ$ glide-reflection boundary. The non-symmorphic symmetry along the 60$^\circ$ glide-reflection boundary can realize the partner change along the invariant line and forces the formation of one-dimensional hourglass~\cite{wang2016hourglass} type metallic GB states as shown in Fig.~\ref{Fig4}c. Since the non-symmorphic symmetry along the GB guarantees the band crossing with a linear dispersion at the low energy~\cite{young2015dirac}, the spectroscopic signature in Fig.~\ref{Fig3}b and ~\ref{Fig4}b looks quite similar with one shown in the edge of topological insulating MoTe$_2$. Contrary to this case, however, the symmorphic symmetry along the 120$^\circ$ two-fold rotation boundary cannot form the band crossing (Fig.~\ref{Fig4}f). Instead, the topological edge states from the left side of the GB interact with those from the right side resulting in gapped boundary states as shown in Fig.~\ref{Fig4}f. So, the band edges in the boundary states make two peaks below and above the charge neutral energy shown in Figs.~\ref{Fig3}f and ~\ref{Fig4}e (Supplementary Section VI).

Our theoretical simulations of STM images and dI/dV spectra agree with experimental observations quite well. In Figs.~\ref{Fig4}a and d, the simulated STM images for the two different GBs structures are displayed. The images of the fully relaxed atomic geometries are computed with the constant current condition~\cite{tersoff1985theory} and shows that the patterns of bright rows of chalcogen atoms clearly respect the underlying crystal symmetries of GBs. We also simulated the local dI/dV spectra right on top of GBs and away from them as shown in Figs.~\ref{Fig4}b and e. The resulting spectra also agree well with our observations showing a single peak at the energetic position of bulk gap for the 60$^\circ$ glide-reflection boundary, while two split peaks for the 120$^\circ$ two-fold rotation boundary. Our projected 1D band structures along the momentum parallel to the grain boundary direction (Figs.~\ref{Fig4}c and f) show sharp differences between two GBs. In Fig.~\ref{Fig4}c for the former structure, the entangled metallic bands fill the band gap of the bulk MoTe$_2$, while the split conduction and valence band states inside the gap are shown in Fig.~\ref{Fig4}f for the latter. So, it is evident that these contrast features of the band structures are responsible for the different dI/dV spectra for the two disparate GBs.

Considering that the typical growth method for the TMDs~\cite{chhowalla2013chemistry}, the GB between crystalline domains with different orientations are inevitable. Thus, our experimental demonstration of protected metallic states here not only provides the first direct evidence of existence of the symmetry enforced Dirac type metallic state along the dislocated atomic defects with the distinct crystal symmetry but also shows a possible route to draw the robust metallic nanowires inside the insulator once the growth of crystalline structure of TMDs can be controlled.

\subsection{Sample preparation}
\label{Sample preparation}
High quality 1T'-MoTe$_2$ single crystals were synthesized using NaCl-Flux method. Stoichiometric mixture of Mo and Te powders were sintered with sodium chloride (NaCl) at 1373~K for 30 hours in evacuated silica tubes. Then, the samples were cooled to 1223~K with a rate of 0.5~K/hr. Rapid cooling down to room temperature was achieved by quenching with water. 1T'-MoTe$_2$ single crystals for this study have large magnetoresistance and residual resistance ratio, exceeding 32,000 and 350, respectively, indicating that Te deficiency is less than 1~\%~\cite{cho2017te}.  

\subsection{Scanning tunneling microscopy/spectroscopy measurements}
\label{Scanning tunneling microscopy/spectroscopy measurements}
We performed the experiments in a commercial low-temperature STM (UNISOKU Co., Ltd., Osaka, Japan) at 2.8~K. 1T'-MoTe$_2$ single crystal sample was cleaved in an ultrahigh vacuum chamber (~10$^{-10}$ Torr) at room temperature and then transferred to the low-temperature STM sample stage, where the temperature was kept at 2.8~K. The STS measurements were performed using a standard lock-in technique with a bias modulation of 5~mV at 1~kHz.

\subsection{Density functional theory and tight-binding calculations}
\label{Density functional theory and tight-binding calculations}
To investigate the structural and electronic properties of 1T'-MoTe$_2$ and their grain boundaries, we performed \textit{ab initio} calculations based on density functional theory (DFT)~\cite{{hohenberg1964phys},{kohn1965phys}} as implemented in VASP code.~\cite{{kresse1996efficient},{kresse1993ab}} Projector augmented wave potentials~\cite{{1994projector},{kresse1999fromultrasoft}} was employed to describe the valence electrons, and the electronic wave functions were expanded by a plane wave basis set with the cutoff energy of 450~eV, and the atomic relaxation was continued until the Hellmann-Feynman force acting on every atom became lower than 0.03~eV/{\AA} . For more precise calculations, we included the dipole correction. The Perdew-Burke-Ernzerhof (PBE) form~\cite{{perdew1996phys},{artacho1999linearscaling}} was employed for the exchange-correlation functional in the generalized gradient approximation (GGA). The Brillouin zone (BZ) was sampled using a $10{\times}10{\times}1$ $k$-grid for the primitive unit cell of 1T' MoTe$_2$. The spin-orbit coupling (SOC) effect and on-site Coulomb repulsion (U) are included in all calculations. These parameters have been thoroughly tested to describe the exact characteristic in the 1T' MoTe$_2$ single layer~\cite{kim2017origins}. For the optimization of GB structures of 1T' MoTe$_2$, we adopted the basis consisting of pseudo-atomic orbitals (PAOs) with a single-$\zeta$ basis set, as implemented in the SIESTA code~\cite{{artacho2008siesta},{portal1997densityfunctional}}. The XC functional was treated with GGA as used for VASP. The behavior of valence electrons was described by a norm-conserving Troullier-Martins pseudopotential~\cite{troullier1993efficient} with scalar-relativistic effect in the Kleinman-Bylander factorized form~\cite{kleinman1982efficacious} $4d5$ $5s1$ $5p1$ and $5s2$ $5p4$ electrons for molybdenum and tellurium are considered as a valence level in pseudopotential. The charge density has been determined self-consistently on a real space mesh with a very high cutoff energy of 450~Ry, sufficient for total energy convergence to within 1~meV/atom. The energy shift due to the spatial confinement of the PAO basis functions~\cite{artacho1999linearscaling} was limited to less than 0.01~Ry. We used periodic boundary conditions and periodic array of slabs separated by a vacuum region of $\geq$ 30~{\AA} throughout the study. The BZ was sampled using a $1{\times}5{\times}1$ and $1{\times}3{\times}1$ $k$-grids for 120$^\circ$ and 60$^\circ$ GBs of 1T'-MoTe$_2$, respectively. To investigate electronic properties of the GBs, we started by constructing Slater-Koster type tight-binding (TB) model which successfully reproduces the DFT band structure for the monolayer 1T-MoTe$_2$ near the Fermi level. Here, we assumed five $d$ orbitals on each Mo atom and $s$, and three $p$ orbitals on each Te atom. Further details of the tight-binding Hamiltonian $H$ in real space are in Supplementary Section VII.

\newpage

\beginsupplement

\subsection{Supplementary Section I. Dependence on the number of layers of 1T'-MoTe$_2$}

\begin{figure}[h]
\includegraphics[width=1.0\columnwidth]{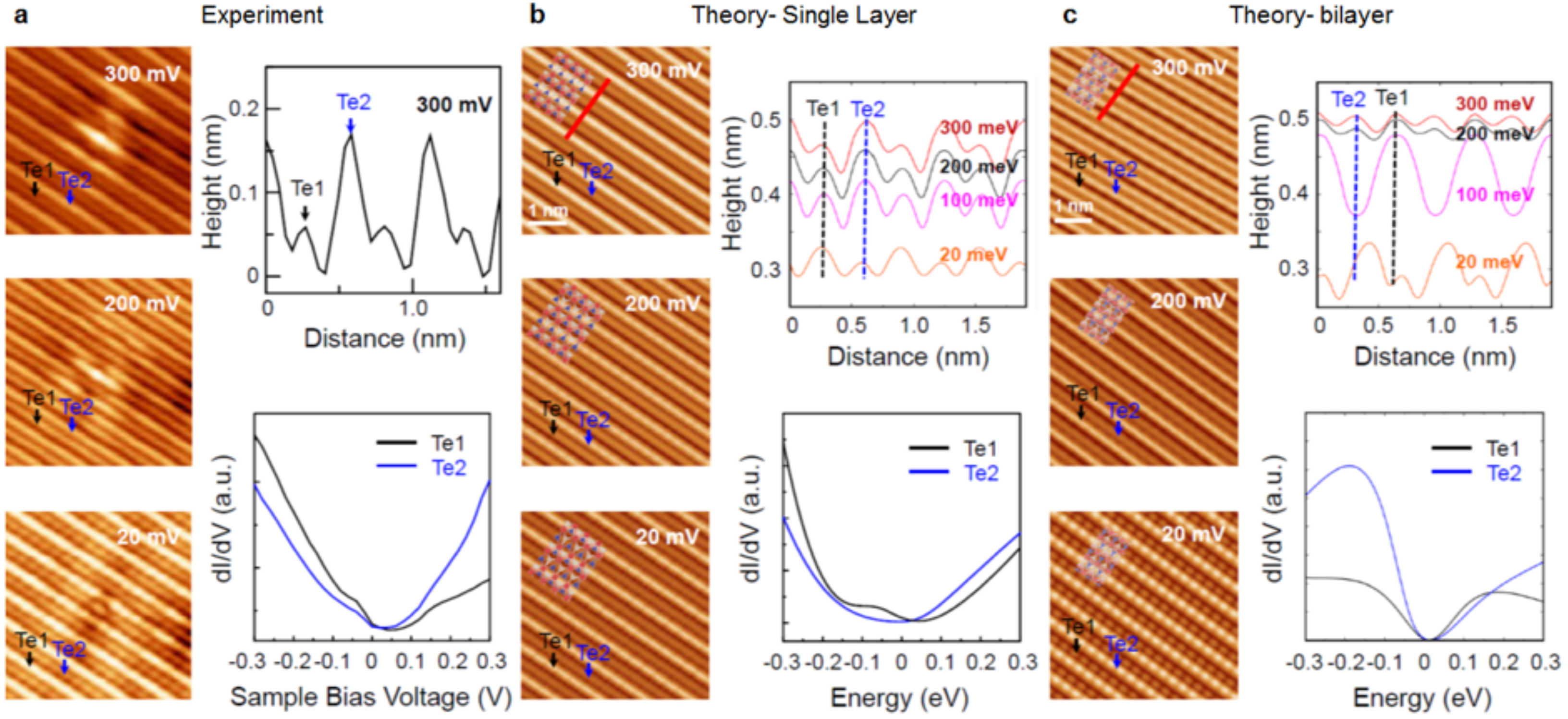}
\caption{Layer thickness dependence of 1T'-MoTe$_2$. a, Experimental STM results. b, c, DFT-calculated results of a single and double layer, respectively. STM images were obtained at 300, 200 and 20~mV for both experiments and calculations. Experimentally obtained dI/dV spectra are quite similar to that of the single-layer rather than the double-layer. Resulting different simulated dI/dV characteristics of the single and double layers of 1T'-MoTe$_2$, the heights changes of Te1 and Te2 rows are differently varied depending applied bias voltages. In the double layer, Te1 is higher than Te2 at 100, 200 and 300~mV. }
\end{figure}

\subsection{Supplementary Section II. Topologically protected metallic state of 1T'-MoTe$_2$.}

\begin{figure}[h]
\includegraphics[width=1.0\columnwidth]{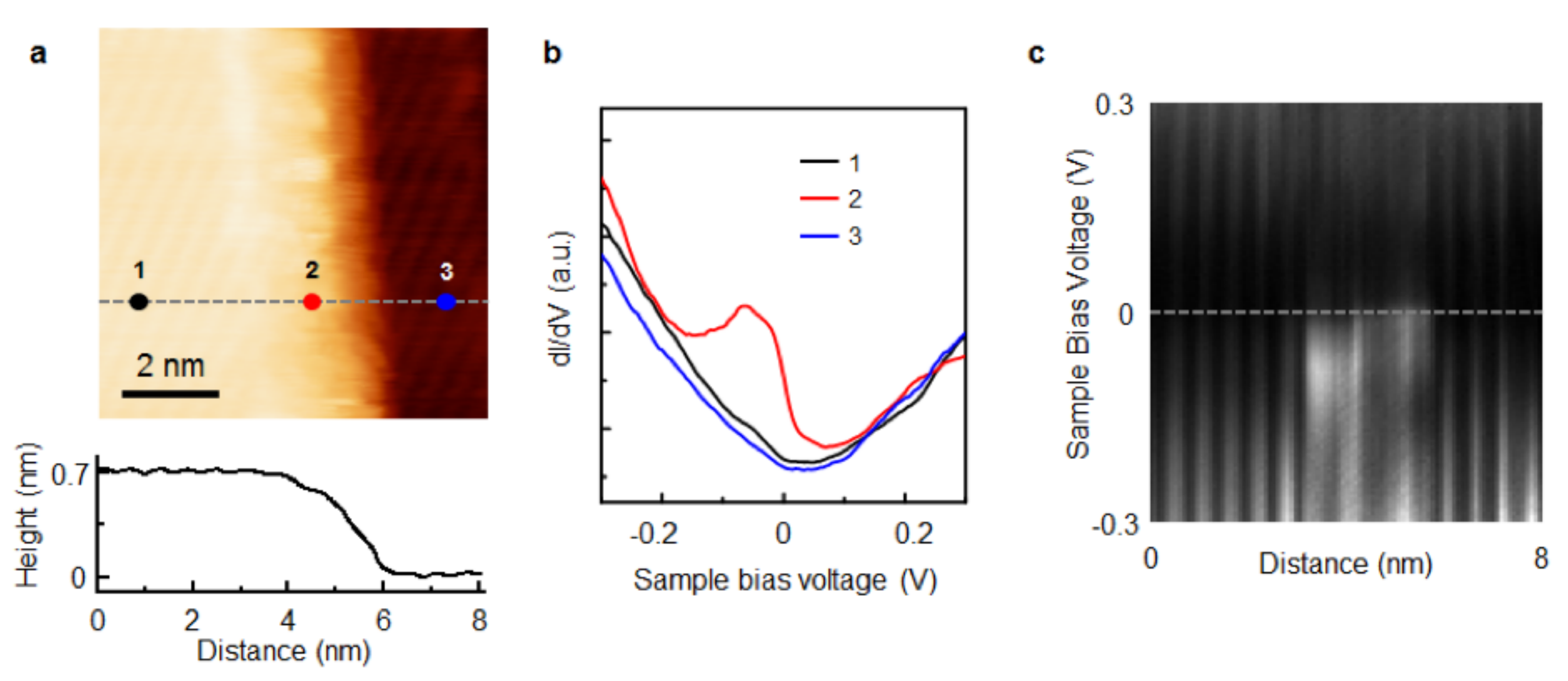}
\caption{Topological edge states at the edge of 1T'-MoTe$_2$. a. STM topography of 1T'-MoTe$_2$ (Vs = 0.3~V, I = 0.1~nA) and profile of the height along the dashed gray line. b. dI/dV spectra obtained at the positions indicated by black, red and blue dots in a. c. dI/dV spectra taken across the step edge of a 1T'-MoTe$_2$.
}
\end{figure}

\newpage

\subsection{Supplementary Section III. Bias-dependence of STM topographs and dI/dV maps of 1T'-MoTe$_2$.}

\begin{figure}[h]
\includegraphics[width=1.0\columnwidth]{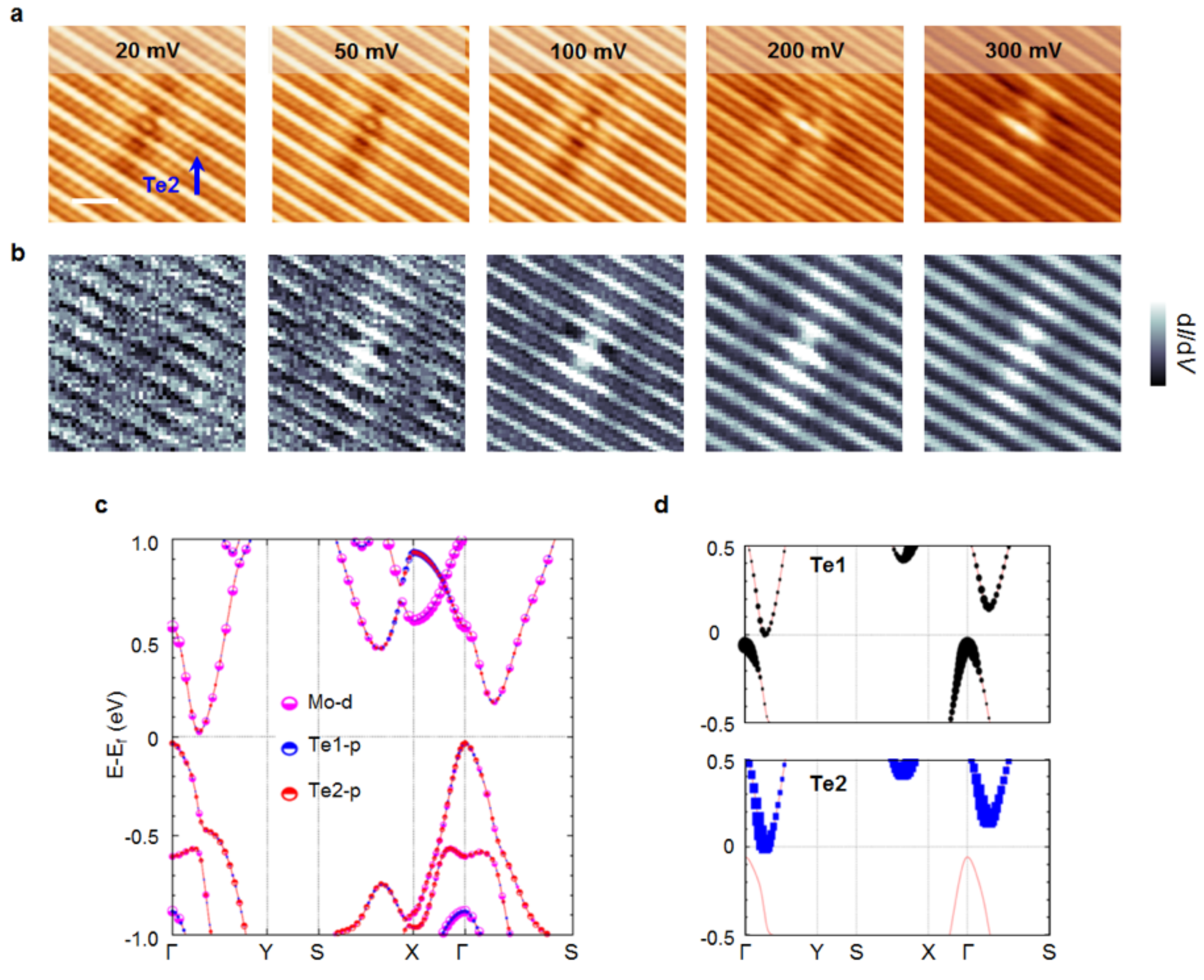}
\caption{Bias-dependence of STM topographs and dI/dV maps of 1T'-MoTe$_2$. a, STM topographs and b, dI/dV maps for sample bias voltages V = 20, 50, 100, 200 and 300 from left to right. Scale bar is 1~nm. A defect is in Te2, which is used for a marker. In the STM images obtained at 20, 50 and 100~mV Te1 is brighter than Te2. At 200~mV the contrast of Te1 and Te2 is similar, and at 300~mV Te2 is brighter than Te1. In the dI/dV maps at 50, 100, 200 and 300~mV Te2 is brighter than Te1. c, Band structure of monolayer 1T'-MoTe$_2$, d, Contributions of Te1 and Te2 p orbitals plotted on the band structure. 
}
\end{figure}

\subsection{Supplementary Section IV. Symmetry and formation energy of grain boundaries}

\begin{figure}[h]
\includegraphics[width=1.0\columnwidth]{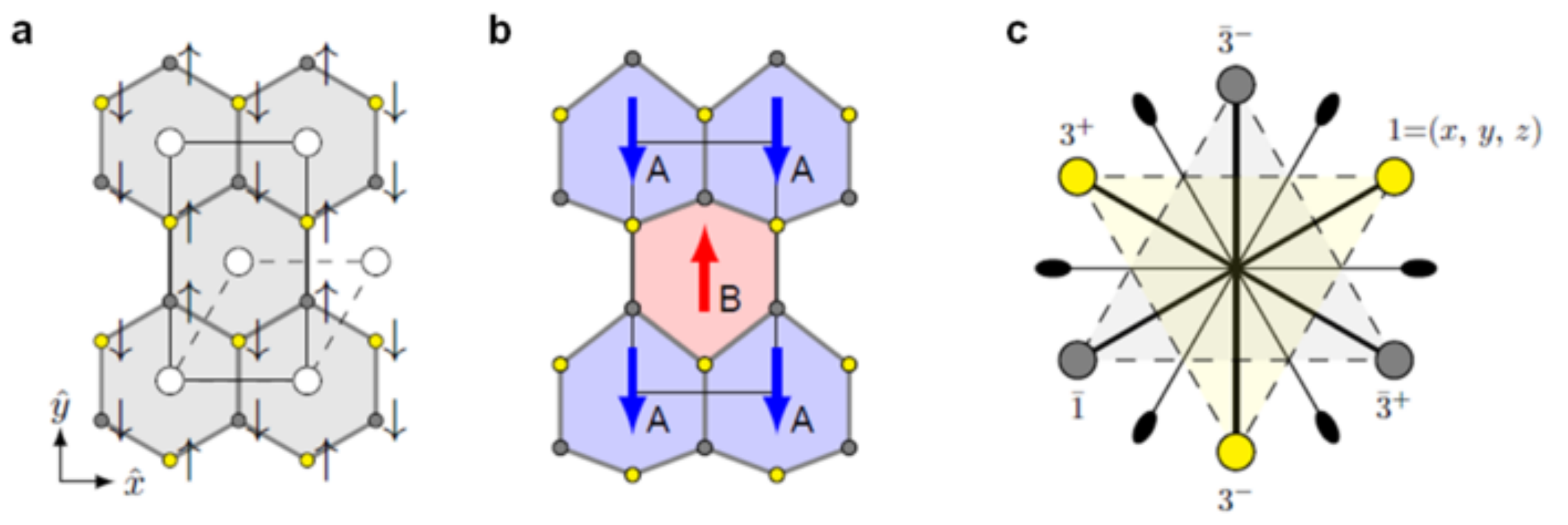}
\caption{Structure of the MoTe$_2$. Atomic structures of the MoTe$_2$ crystal in a, T (P-3m1) and b, T' (P2$_1$/m) phases. The large white balls represent Mo atoms, and Te atoms are marked as the smaller balls. A phase transition from T to T' phase is indicated by the arrows along $\pm y$ directions next to the Te atoms. The primitive unitcell for the T phase is shown in dashed lines in a. c, A point group element of $-3$. There exist three diads (two-fold rotation axes) along one of the edges of the triangles shaded in gray and yellow, and three mirror lines (thick lines) dissecting one of the edges.
}
\end{figure}

Here, we demonstrate a grain boundary (GB) model to provide atomic models of the observed GBs. The model provides possible atomic rearrangements at the GB based on coincidence site lattice (CSL) theory and point group analysis, especially when the given crystal is stabilized by lowering its symmetry, e.g., Jahn-Teller distortion. Instead of constructing a GB directly from the original crystal which has lower symmetry, we generate GB models in two-step process. Firstly, we create a bi-crystal separated by a symmetric tilt GB from high-symmetry intermediate crystal by using CSL theory. The intermediate crystal can be found by searching for the best-matched-fit molecular geometry from the original crystal with the lower symmetry~\cite{supp1}. After that, the intermediate crystals in each domain are distorted back to the original crystal.

\begin{figure}[h]
\includegraphics[width=0.7\columnwidth]{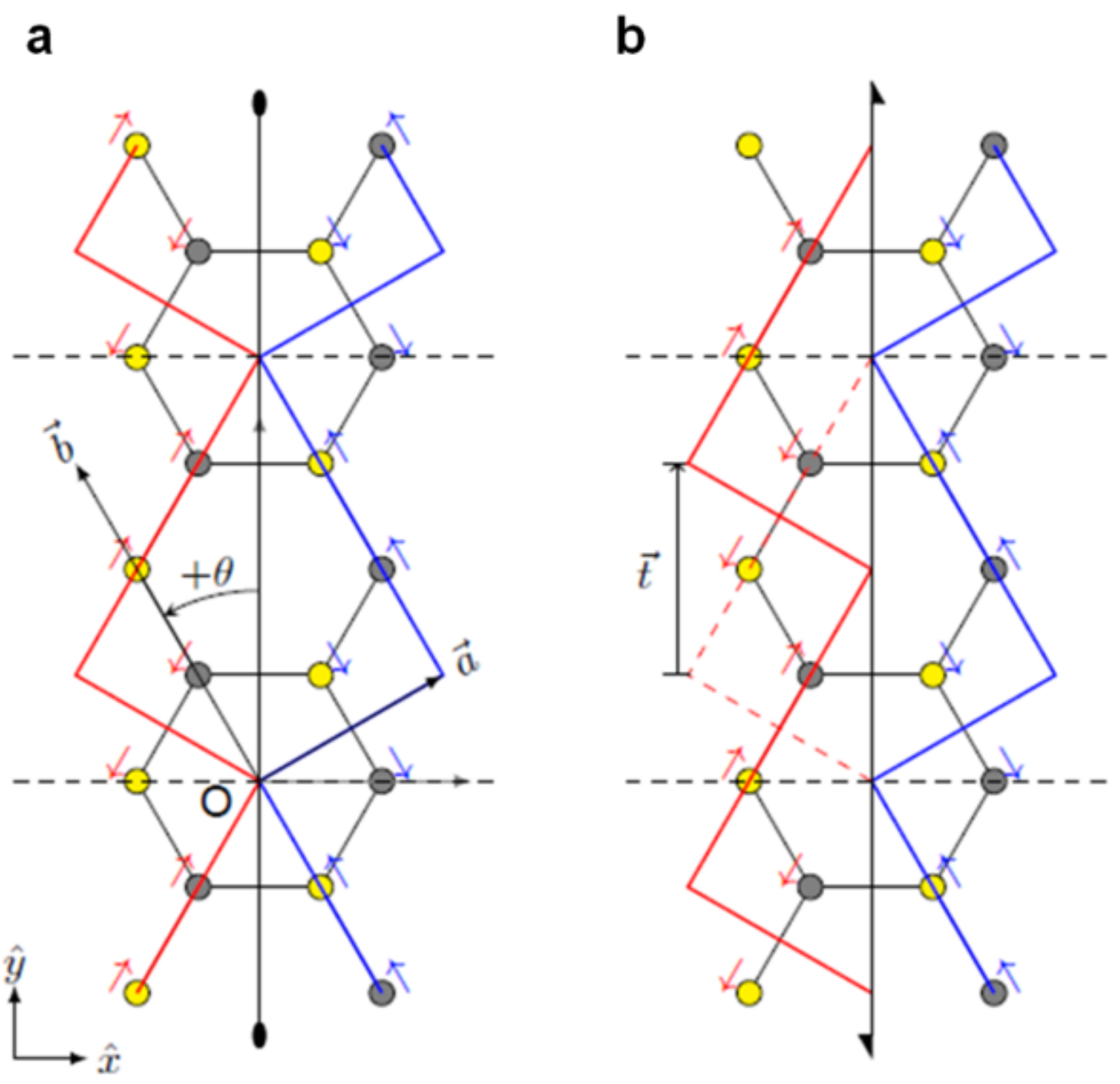}
\caption{Structural units near the two-fold symmetric grain boundaries with a, symmorphic and b, non-symmorphic symmetries.}
\end{figure}

\begin{figure}[h]
\includegraphics[width=0.7\columnwidth]{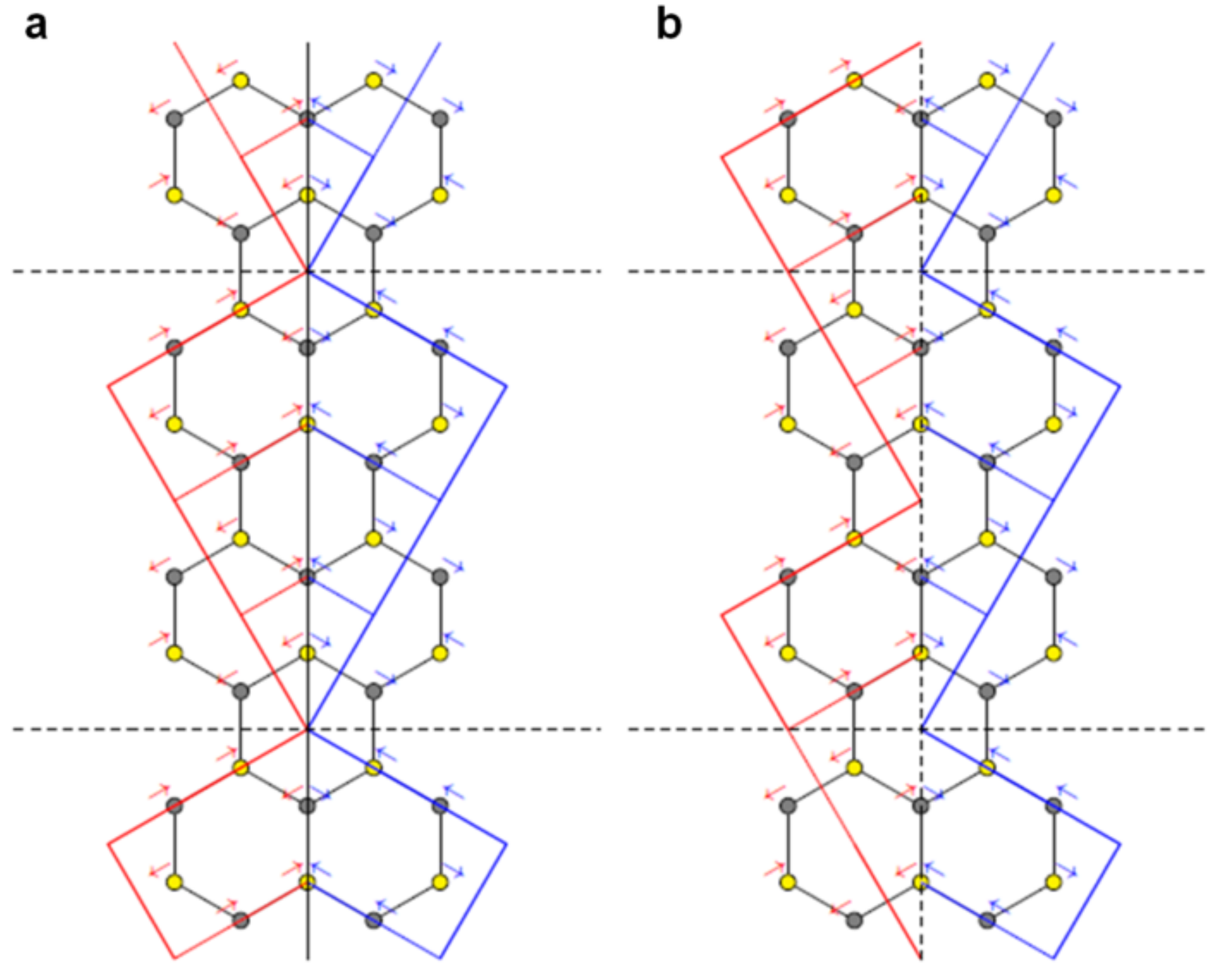}
\caption{Structural units near the mirror-symmetric grain boundaries with a, symmorphic and b, non-symmorphic symmetries. }
\end{figure}

In the case of a monolayer MoTe$_2$, a distorted octahedral motif in a T' phase can be best-matched to the regular octahedron with a point group symmetry of $-3$ (or six-fold improper rotation: S$_6$ by Schoenflies notation), and the space group of the crystal is changed from P2$_1$/m (no. 11) phase to P-3m1 (no. 164) (T phase) as shown in Figs. S4a and b. In addition to the improper rotation, there are three additional two-fold rotation axes perpendicular to the improper rotation axis and three vertical mirror planes (Fig. S4c). This indicates that a GB, of which relative angle of rotation ($2\theta$) is 60$^\circ$, will create a C$_{2y}$-symmetric interface with the GB angle of $\pi-2\theta$, i.e., 120$^\circ$ (Fig. S5). Similarly, $2\theta$ of 120$^\circ$ results in M$_y$-symmetric interface with the GB angle of 60$^\circ$ (Fig. S6). The above also hold when fractional translation is followed, and both of the ``symmorphic" and ``non-symmorphic" GBs are energetically degenerated prior to the distortion. Once the distortion is taken into account as indicated by arrows in Figs. S5 and S6, the energies of symmorphic and non-symmorphic GBs become different, and the four GB models as in Figs. 2b-e are constructed. Since each of the models is tractable by GB operations by which one side of the bi-crystal can be mapped onto the other, we label each of the GB models as follows: 60$^\circ$ mirror $\{M_y |0\}$, 60$^\circ$ glide-reflection $\{M_y |t\}$, 120$^\circ$ two-fold rotation $\{C_{2y}|0\}$ and 120$^\circ$ screw $\{C_{2y} |t\}$, respectively.

\subsection{Supplementary Section V. Differential conductance maps}

\begin{figure}[h]
\includegraphics[width=1.0\columnwidth]{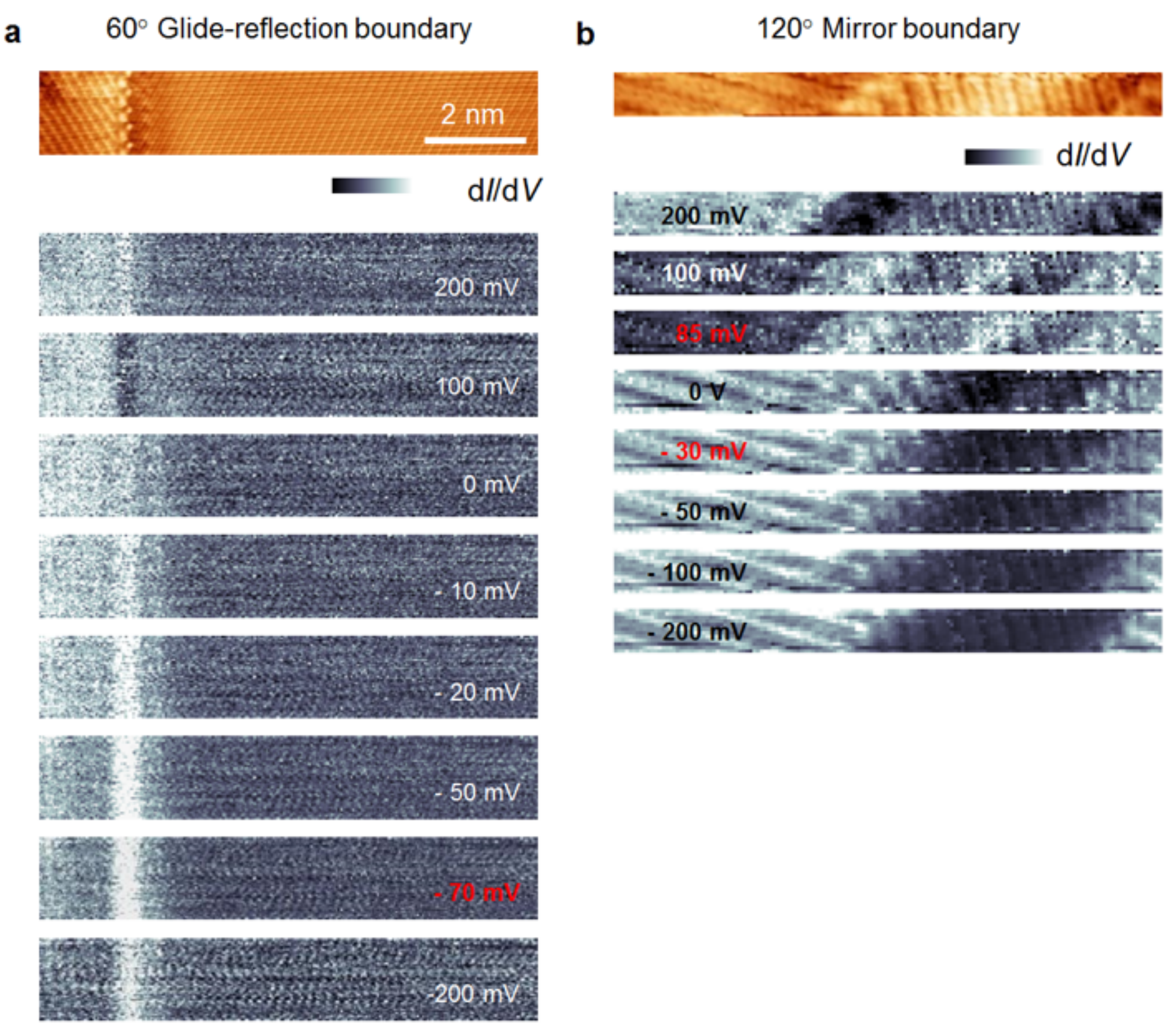}
\caption{Differential conductance maps. a. STM topograph and dI/dV maps of 60$^\circ$ glide-reflection boundary for bias voltages V = 200, 100, 0, $-$10, $-$20, $-$50, $-$70 and $-$200~mV from top to bottom. b. STM topograph and dI/dV maps of 120$^\circ$ mirror boundary for bias voltages V = 200, 100, 85, 0, $-$30, $-$50, $-$100 and $-$200~mV from top to bottom }
\end{figure}

\subsection{Supplementary Section VI. DFT-calculated DOS of Te atoms in 120$^\circ$ two-fold rotation boundary}

\begin{figure}[h]
\includegraphics[width=1.0\columnwidth]{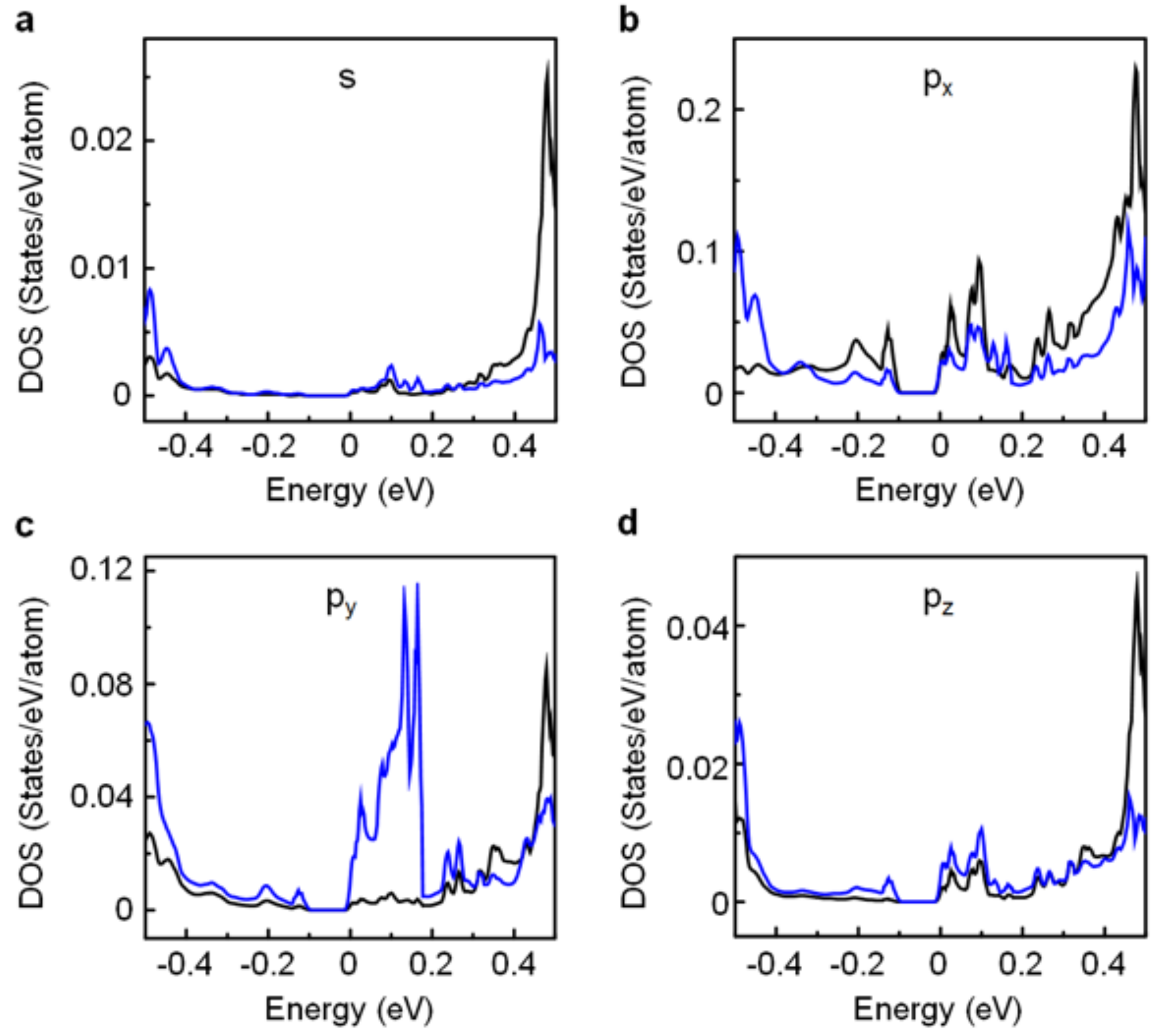}
\caption{DFT-calculated DOS projected on $s$, $p_{x}$, $p_{y}$ and $p_{z}$ orbitals of Te atoms in 120$^\circ$ two-fold rotation boundary. }
\end{figure}

\subsection{Supplementary Section VII. Tight-binding model for 1T'-MoTe$_2$}

To investigate electronic properties of the grain boundaries, we start by constructing Slater-Koster type tight-binding (TB) model which successfully reproduces the DFT band structure for the monolayer 1T'-MoTe$_2$ near the Fermi level. Here, we assume five $d$-orbitals on each Mo atom and $s-$, and three $p$-orbitals on each Te atom. The tight-binding Hamiltonian $H$ in real space is given as follows:
$$
H=\sum_{\langle i,j\rangle}\sum_{\sigma\alpha\alpha'}\left[t^{\alpha\alpha'}_{i,j}c^\dagger_i c_j+\textrm{c.c.}\right]
+H_\textrm{soc},
$$
where $i(j)$ labels the atomic sites, $\sigma$ spin and $\alpha(\alpha')$ orbitals, and $t^{\alpha\alpha'}_{i,j}$ is a transfer matrix, which can be parameterized depending on the direction and distance between pair of orbitals through the Slater-Koster formula~\cite{supp2}. The $H_\textrm{soc}$ represents the effect of on-site spin-orbit coupling (SOC),
$$
H_\text{soc}=-\lambda_\textrm{Mo}\hat{S}\cdot\hat{L}_\textrm{Mo}-\lambda_\textrm{Te}\hat{S}\cdot\hat{L}_\textrm{Te}
$$
where $\lambda_\textrm{Mo(Te)}$ are SOC parameters for Mo (Te) atom, $\hat{S}$ is the spin 1/2 operator, and 
$\hat{L}_\textrm{Mo(Te)}$ is the angular momentum operator of Mo (Te) atom, respectively~\cite{supp3}. Then, TB parameters are fitted by minimizing fitness function $F$,
$$
F=\sum_{n,{\bf k}}\omega_{n{\bf k}}(E_{n{\bf k}}^\textrm{TB}-E_{n{\bf k}}^\textrm{DFT})^2
$$
where $\omega_{n{\bf k}}$ is weight at the k-point and $n$-th eigenvalue, and $E_{n{\bf k}}^\textrm{TB(DFT)}$ is eigenvalue obtained by TB (DFT) for monolayer 1T'-MoTe$_2$. We use a nonlinear least-squares method bases on the Levenberg-Marquardt algorithm~\cite{supp4,supp5}. With this procedure, we successfully fit the DFT band structure for 1T'-MoTe$_2$, its topological properties and the orbital character compared with those from the DFT calculations.

\end{document}